\documentclass[a4paper,11pt]{article}
\pdfoutput=1 

\usepackage{jinstpub} 
\usepackage{graphicx}
\usepackage{float}
\usepackage{caption}
\usepackage{array}
\usepackage{subfigure}
\usepackage{multirow}
\usepackage{siunitx}
\usepackage{xfrac}
\usepackage{verbatim}
\usepackage{pifont}
\usepackage{arydshln}
\usepackage{paralist}
\usepackage{enumitem}
\usepackage{layout} 
\usepackage{amsmath}
\usepackage{csquotes} 
\usepackage{rotating}
\usepackage{adjustbox}
\usepackage{feynmp}
\DeclareSIUnit\torr{Torr}
\DeclareSIUnit\cee{c}
\DeclareSIUnit\electronvoltnr{eV_r}
\usepackage[T1]{fontenc}
\usepackage[utf8]{inputenc}

\title{\boldmath Demonstration of radon removal from SF$_6$ using molecular sieves}

\author{A.~C.~Ezeribe}
\author{, W.~Lynch}
\author[1]{, R.~R.~ Marcelo Gregorio\note{Corresponding author.}} 
\author{, J.~ Mckeand}
\author{, A.~ Scarff }
\author{and N.~J.~C.~Spooner}

\affiliation{Department of Physics and Astronomy, University of Sheffield, S3 7RH, U.K.}

\emailAdd{mr.rgregorio@live.com}

\abstract{The gas SF$_6$ has become of interest as a negative ion drift gas for use in directional dark matter searches.  However, as for other targets in such searches, it is important that radon contamination can be removed as this provides a  source of unwanted background events.  In this work we demonstrate for the first time filtration of radon from SF$_6$ gas by using a molecular sieve.  Four types of sieves from Sigma-Aldrich were investigated, namely 3{\AA}, 4{\AA}, 5{\AA} and 13X.  A manufactured radon source was used for the tests.  This was attached to a closed loop system in which gas was flowed through the filters and a specially adapted Durridge RAD7 radon detector. In these measurements, it was found that only the 5{\AA} type was able to significantly reduce the radon concentration without absorbing the SF$_6$ gas. The sieve was able to reduce the initial radon concentration of 3875 $\pm$ 13 Bqm$^{-3}$ in SF$_6$ gas by 87\% when cooled with dry ice. The ability of the cooled 5{\AA} molecular sieve filter to significantly reduce radon concentration from SF$_6$ provides a promising foundation for the construction of a radon filtration setup for future ultra-sensitive SF$_6$ gas rare-event physics experiments.}


\begin{document}
\maketitle
\flushbottom

\section{Introduction}\label{sec:intro}
Radon contamination is a problem in ultra-sensitive gas rare-event physics experiments such as DRIFT  (Directional Recoil Identification From Tracks) \cite{sadler2014} because the decay of radon gas inside the detector can be a source of events able to mimic genuine signals. Therefore, minimising radon concentration in these experiments is important. In DRIFT and other detectors, this is achieved in part by continuous flow and disposal of the target gas.

 SF$_6$ is a candidate to replace current target gas in directional rare-event search experiments due to its novel properties \cite{phan2017}. However, SF$_6$ is a potent greenhouse gas; for the same mass of gas, SF$_6$ traps heat nearly 24,000 times more than CO$_2$ \cite{solomon2007}. This makes disposing of SF$_6$ problematic. One alternative is to introduce continuous recirculation and reuse of the SF$_6$ gas with active removal of radon using an appropriate filtration process.  This is to reduce the radon contamination level of the target SF$_6$ gas to the required $\sim$\si{\micro\becquerel\per\litre} range. The source of this radon is the alpha decay of radium-226 in the decay chain of natural Uranium isotopes in the gas and materials of the gas cylinder.
  
Reported herein is the filtration of radon from SF$_6$ gas by using molecular sieves. Initially, this was done by determining the types of molecular sieves that do not absorb SF$_6$. This was followed by an investigation on the ability of the molecular sieve types to reduce the radon concentration in the contaminated SF$_6$ gas. Finally, optimisation of the molecular sieve adsorption was explored by the application of a cold trap.

\newpage

\section{Molecular Sieve}\label{Molecularsieve}
\begin{figure}[h]
\begin{center}
\includegraphics[scale=0.35]{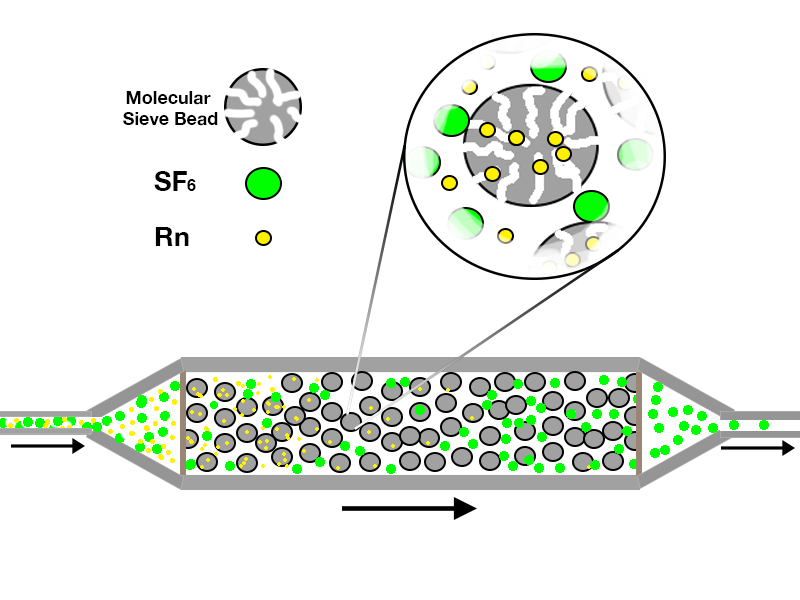} 
\vspace*{-10mm}
\caption{Illustration of the separation of smaller radon atom from larger SF$_6$ molecules using molecular sieves. The arrows represent the direction of flow of the gas inside the molecular sieve vessel.}
\label{molsiv}
\end{center}
\end{figure}

\noindent Molecular sieves are crystalline metal aluminosilicates structures with specific pore sizes \cite{molecularsieves}. These pores allow molecules with the critical diameter equal or below the pore size to diffuse and be adsorbed on to the structure; but it allows molecules with diameters larger than the critical diameters to pass between the bead gaps. This process is illustrated in Figure \ref{molsiv}.\\

\noindent The range of the diameter of radon is 2.5{\AA} - 4.9{\AA}  \cite{ghosh2011,alvarez2008,rahm2016,Chen2014} depending on how the diameter is defined. For example, it can be defined as the van dar Waal, covalent or Ionic radii. Due to the different definitions, this range of diameter of radon does not necessarily equal to the critical diameter appropriate for any of the molecular sieves. Thus, for this work all appropriate available types of molecular sieves were considered; 3{\AA}, 4{\AA}, 5{\AA} and 13X. The number before the {\AA} in the first three named sieves corresponds to the sieve pore size in {\AA}ngstr$\ddot{\text{o}}$m, whereas for the 13X type the pore size is 10 {\AA}ngstr$\ddot{\text{o}}$m.

\section{Experimental Setup}

\begin{figure}[t!]
\begin{center}
\includegraphics[scale=0.8]{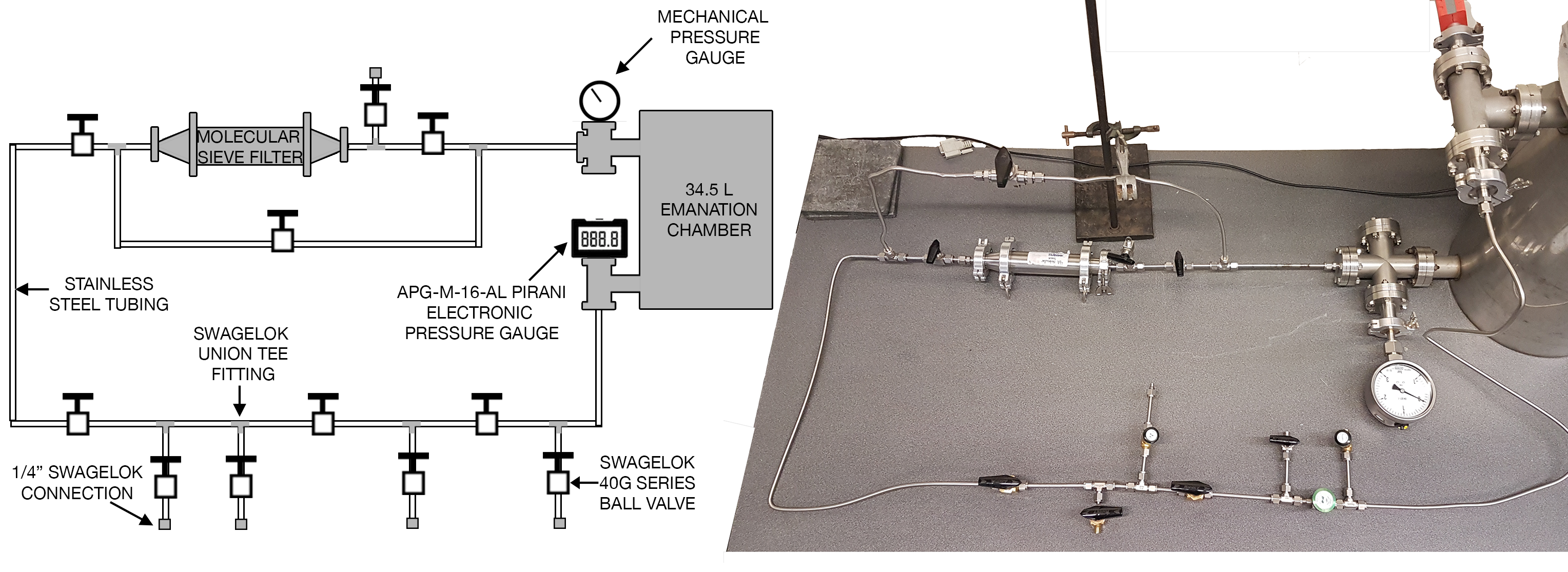} 
\caption{A schematic of the constructed setup used for calibration of the molecular sieves (left panel). In the right panel is a picture of the calibration setup.}
\label{real}
\end{center}
\end{figure}

To determine the molecular sieve with the appropriate critical diameter for adsorption of radon from SF$_6$, the experimental setup shown in Figure \ref{real} was used. This comprises a loop of stainless steel pipes allowing SF$_6$ gas to be circulated through an emanation chamber and the molecular sieve filter. The emanation chamber was used to enhance radon-SF$_6$ equilibrium. This chamber was attached to a digital and analogue pressure gauge for complementary pressure measurements. Swagelok union tees and quarter turn valves fittings were used to create junctions before and after the molecular sieve filter so it could be engaged and disengaged when required. In addition, various inlet and outlet valves were connected throughout the loop.\\ 

\begin{figure}[h]
\begin{center}
\includegraphics[scale=0.2]{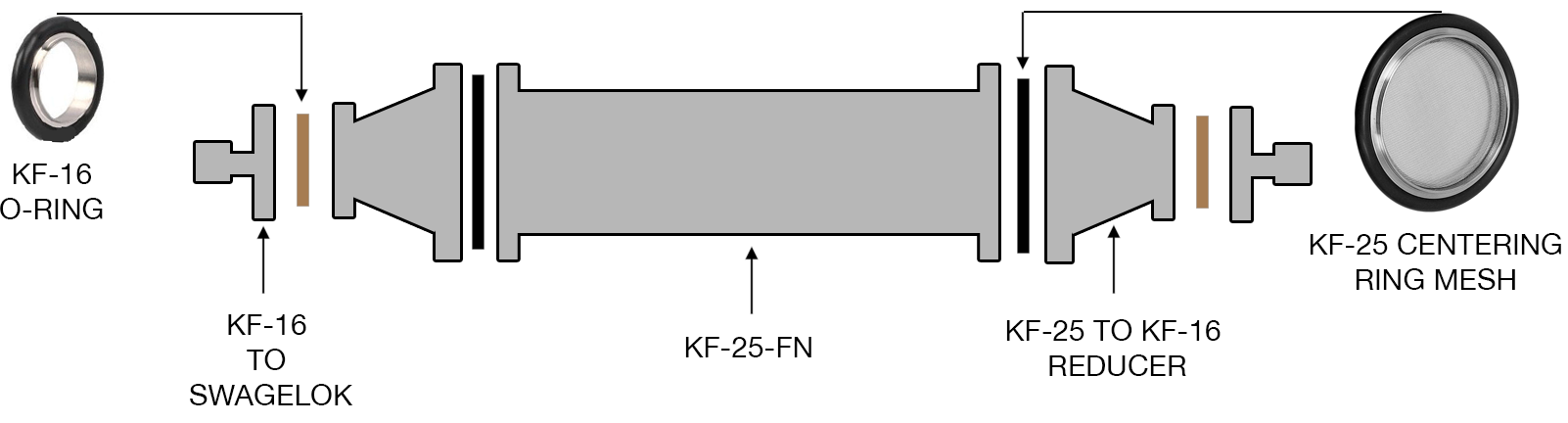}
\vspace*{-3mm}
\caption{Diagram of the molecular sieve filter with detailed components.}
\label{msf}
\end{center}
\end{figure}

\noindent The molecular sieve beads were contained inside a KF-25-FN vessel using KF-25 centering O-ring meshes as shown in Figure \ref{msf}.  The vessel holds up to 100g of each type of the molecular sieve beads. Details of the Sigma-Aldrich molecular sieves examined are shown in Table \ref{MSFtypesss}. The pore size shown corresponds to the size of the cavities in the molecular sieve structure. An atom or molecule with critical diameter lower than the pore size is trapped within the cavity. The molecular formula arises from the process of synthesising the metal aluminosilicate structure which heavily depends on the species substituted to create the cavity.

\begin{table}[h]
\centering
\scalebox{0.7}{\begin{tabular}{|c|ccc|}
\hline
\textbf{\begin{tabular}[c]{@{}c@{}}Molecular \\ Sieve\end{tabular}} & \textbf{Molecular Formula}                         & \textbf{\begin{tabular}[c]{@{}c@{}}Pore Size \\ ({\AA}ngstr$\ddot{\text{o}}$ms)\end{tabular}} & \textbf{\begin{tabular}[c]{@{}c@{}}Approx. Bead \\ Size (mm)\end{tabular}} \\ \hline
3{\AA}                                                                  & $\rm 0.6K_2O\cdot 0.4 Na_2O \cdot Al_2O_3$             & 3                                                                         & 2                                                                          \\
4{\AA}                                                                  & $\rm Na_2O \cdot Al_2O_3 \cdot 2.0SiO_2$               & 4                                                                         & 2                                                                          \\
5{\AA}                                                                  & $\rm 0.80CaO\cdot 0.20Na_2O \cdot Al_2O_3 \cdot SiO_2$ & 5                                                                         & 4                                                                          \\
13X                                                                 & $\rm Na_2O \cdot Al_2O_3 \cdot 2.8 SiO_2$              & 10                                                                        & 4                                                                          \\ \hline
\end{tabular}}
\caption{Properties and specifications of the molecular sieves that were examined.}
\label{MSFtypesss}
\end{table}

\section{SF$_6$ Absorption Test}
Before testing the radon filtration capabilities of the molecular sieves it was important to verify whether they absorb SF$_6$ molecules or allow them to pass through easily.

\subsection*{Procedure for Testing SF$_6$ Absorption }
To test the absorption of SF$_6$ gas, several components were attached to the constructed testing setup shown in Figure \ref{real}. As shown in Figure \ref{radonpress}, these include: a vacuum scroll pump; used for gas circulation and evacuation, a \textit{Drierite} desiccant connected between the gas canister and the system; used to remove any moisture from the incoming SF$_6$ gas. The minimum purity of SF$_6$ gas used in these measurements is 99.9\si{\percent}.\\

\begin{center}
 \begin{minipage}{\textwidth}
  \begin{minipage}[b]{0.49\textwidth}
    \centering
  \includegraphics[width=\textwidth]{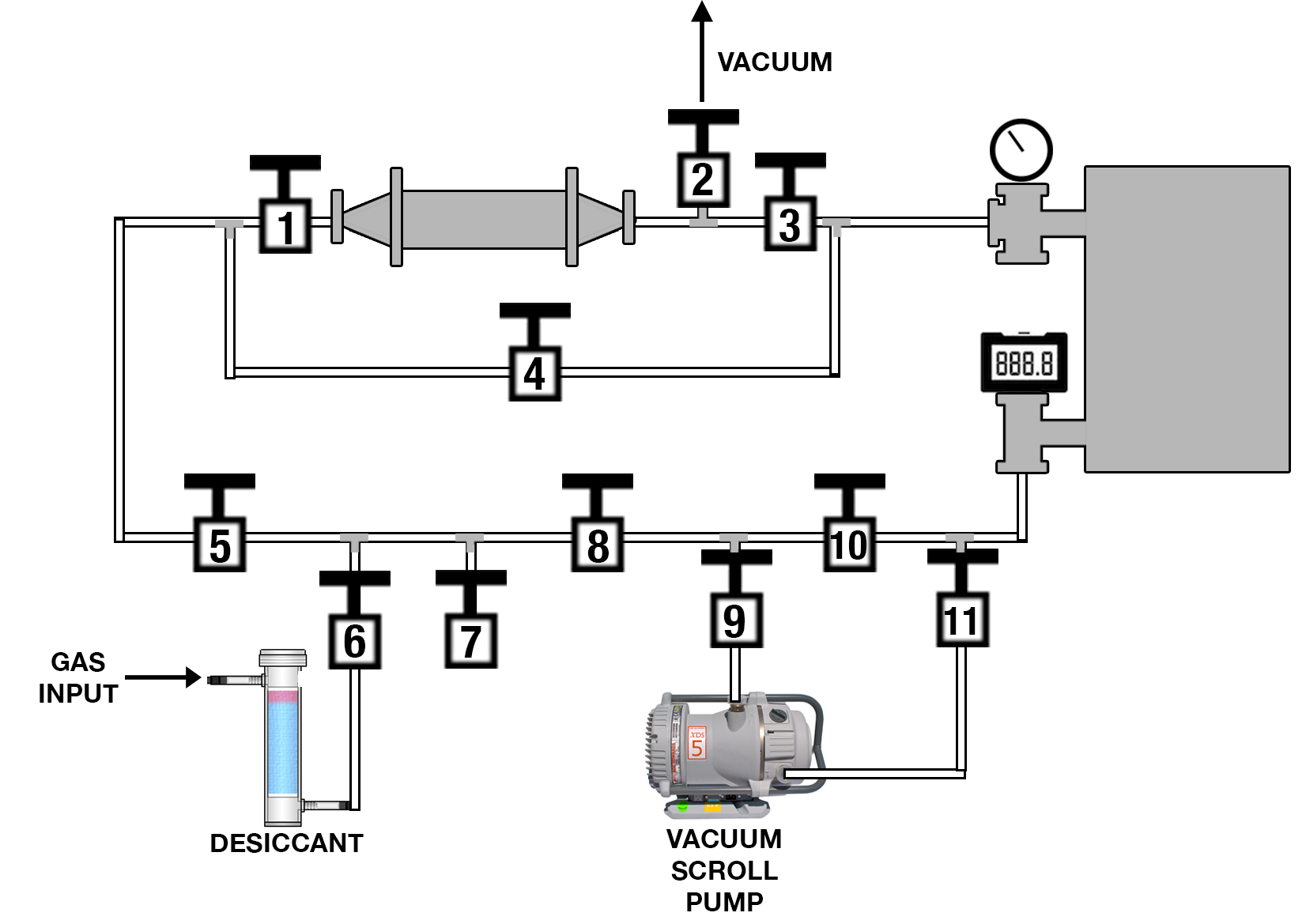} 
    \captionof{figure}{Schematic of the components required for testing the absorption of SF$_6$.  The valves are labelled with numbers.}
    \label{radonpress}
  \end{minipage}
  \hfill
  \begin{minipage}[b]{0.49\textwidth}
    \centering
  \begin{table}[H]
\centering

\scalebox{0.7}{
\begin{tabular}{|c|ccccccccccc|}
\hline
\multirow{2}{*}{\textbf{Step}}                                                      & \multicolumn{11}{c|}{\textbf{Closed Valves}}                                                                                                                                                                                                                                                                                  \\
                                                                                    & \textbf{1}                      & \textbf{2}           & \textbf{3}                      & \textbf{4}           & \textbf{5}           & \textbf{6}                      & \textbf{7}                      & \textbf{8}           & \textbf{9}                      & \textbf{10}          & \textbf{11}                      \\ \hline
Evacuation of System                                                                &                                 &                      &                                 &                      &                      & \textbullet                     & \textbullet                     &                      & \textbullet                     &                      & \textbullet                      \\ \hline
Input of Gas                                                                        &    \textbullet                             & \textbullet          &                                 &                      &                      &                                 & \textbullet                     &                      & \textbullet                     &                      & \textbullet                      \\ \hline
\begin{tabular}[c]{@{}c@{}}Initial Equilibrium\\  Pressure Measurement\end{tabular} & \textbullet                     & \textbullet          &   \textbullet                  &                      &                      & \textbullet                     & \textbullet                     &                      &                                 & \textbullet          &                                  \\ \hline
Engagement of Filter                                                                &                                 & \textbullet          &                                 & \textbullet          &                      & \textbullet                     & \textbullet                     &                      &                                 & \textbullet          &                                  \\ \hline
\begin{tabular}[c]{@{}c@{}}Pressure Measurement\\ Against Time\end{tabular}         &                                 & \textbullet          &                                 & \textbullet          &                      & \textbullet                     & \textbullet                     &                      &                                 & \textbullet          &                                  \\ \hline
Changing of Filter                                                                  & \multicolumn{1}{l}{\textbullet} & \multicolumn{1}{l}{} & \multicolumn{1}{l}{\textbullet} & \multicolumn{1}{l}{} & \multicolumn{1}{l}{} & \multicolumn{1}{l}{\textbullet} & \multicolumn{1}{l}{\textbullet} & \multicolumn{1}{l}{} & \multicolumn{1}{l}{\textbullet} & \multicolumn{1}{l}{} & \multicolumn{1}{l|}{\textbullet} \\ \hline
\end{tabular}}

\end{table}
      \captionof{table}{Steps of the SF$_6$ absorption test operations and state of the valves. Closed valves are marked with solid black dots, unmarked valves were open in each of the operations. The numbers correspond to the valves in Figure \ref{radonpress}.}
      \label{pressvalves}
    \end{minipage}
  \end{minipage}
  
\end{center}

\vspace*{10mm}

\noindent To test whether the molecular sieves absorb SF$_6$, each of the sieve filters were added to the setup and evacuated with the vacuum scroll pump for an hour. It was then filled with SF$_6$ gas via the desiccant up to approximately 40 Torr (typical operational pressure in some directional rare event search experiments). To prevent the absorption of SF$_6$ before the measurement, one of the filter valves was closed during the filling process as shown in Table \ref{pressvalves}. Once the gas was filled to the desired pressure, the filter valves were closed and the vacuum scroll pump was used to create a gas current. When a constant pressure was reached, the molecular sieve filter was engaged and the measurement of the setup's pressure against time using the digital pressure gauge was recorded.

To avoid wasting the gas, the same gas was reused for all the other sizes of sieves tested. Sieve replacement was done by closing the appropriate valves and detaching the filter from the system. Before the new filter was reattached and reintroduced to the system, it was evacuated with the vacuum scroll pump to avoid contamination of the gas.

\subsection*{SF$_6$ Absorption Results}

\begin{figure}[h]
\begin{center}
\includegraphics[scale=0.35]{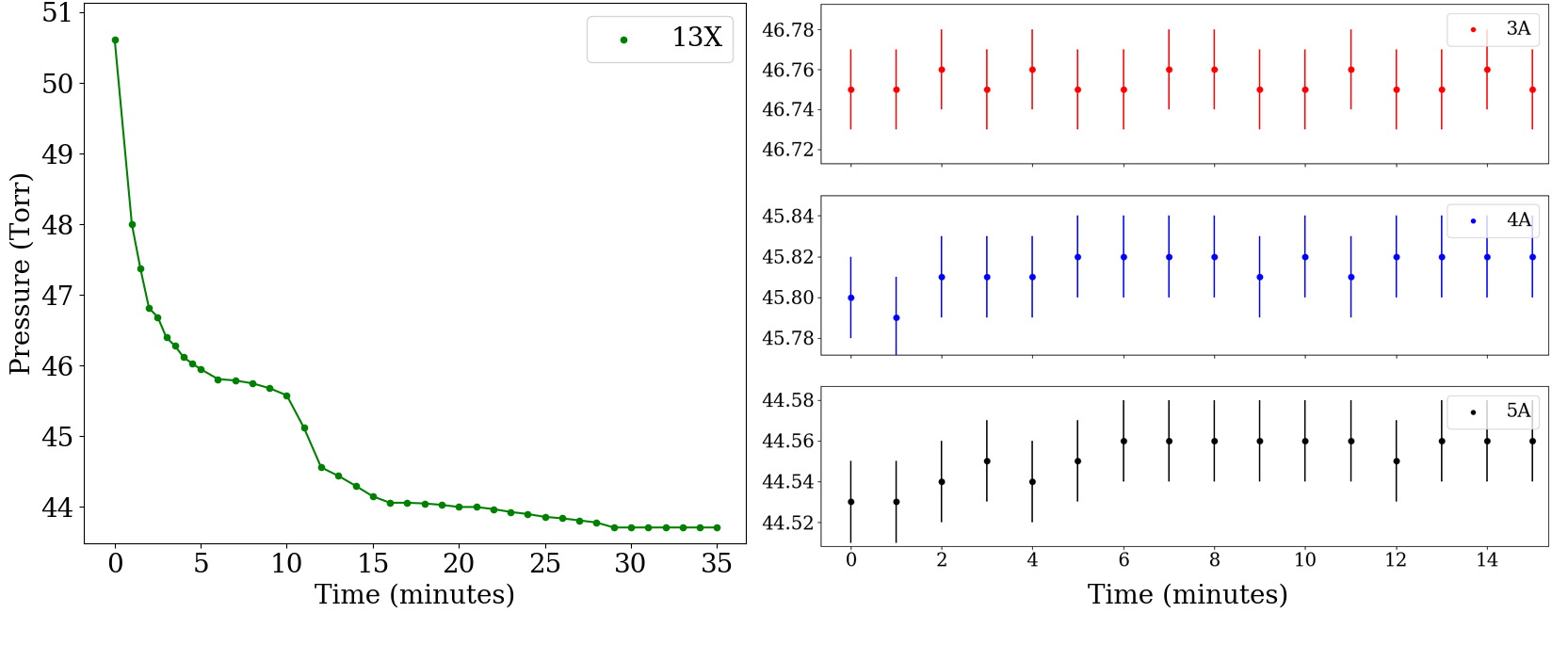}
\caption{The pressure of the system as a function of time for 13X, 3{\AA}, 4{\AA} and 5{\AA} molecular sieve filters. The filters were engaged at time zero for each individual measurement. The errors for the pressure measurement is $\pm$ 0.02 Torr; too small to be seen in the 13X data scale. }
\label{SF6results}

\end{center}
\end{figure}

The pressure change of the system as a function of time for the 13X, 3{\AA}, 4{\AA} and 5{\AA} molecular sieves are shown in Figure \ref{SF6results}. For the 13X molecular sieve, it can be seen that the pressure of the system decreased significantly once the filter was engaged. Pressure measurements were recorded until the sieves were saturated.  It was found that after 35 minutes there was a total pressure decrease of 6.92 $\pm$ 0.02 Torr. Conversely, no significant pressure variation was observed when the 3{\AA}, 4{\AA} and 5{\AA} molecular sieve types were used.

\noindent The significant pressure decrease in the 13X molecular sieve filter test indicates that the sieve absorbs SF$_6$ gas. Therefore, this sieve was not used for further radon filtration tests. This result was expected as the technical information provided by Sigma-Aldrich suggest that the major application of 13X type sieves are with long chain hydrocarbons \cite{molecularsieves}, whereas SF$_6$ is an unchained simple molecule.

The absence of a major pressure change in the 3{\AA}, 4{\AA} and 5{\AA} molecular sieve tests indicates that the SF$_6$ gas is not significantly absorbed. This implies that the critical diameter of SF$_6$ is bigger than 5{\AA} pore size, hence bigger than the pore sizes for 4{\AA} and 3{\AA} types.  Therefore, the radon filtration capabilities of the 3{\AA}, 4{\AA} and 5{\AA} molecular sieves can be tested with confidence that the SF$_6$ gas will pass through the molecular sieve filters. Additionally, results from these measurements constrain the critical diameter of SF$_6$ molecules as 5 - 10 {\AA}ngstr$\ddot{\text{o}}$m.

\newpage
\section{Radon Filtration}\label{sec:rad}
Having established that the 3{\AA}, 4{\AA} and 5{\AA} molecular sieves do not absorb the background gas SF$_6$, the capabilities of each of the sieves to filter radon from the gas were investigated.

\subsection*{Procedure for Testing Filtration of Radon }
To test the ability of molecular sieves to filter radon from SF$_6$ gas samples, a passive 5.361~\si{\kilo\becquerel} radon gas source from Pylon electronics Inc and a RAD7 radon detector \cite{durridge} was added to the setup as shown in Figure \ref{radonfilt}. The source contains a radioactive dry $^{226}$Ra, which decays into radon gas in an aluminium canister.  Using this radon gas, each of the SF$_6$ samples was contaminated for 1 hour by passive diffusion. The RAD7 detector has an internal pump (of rate 1~\si{\litre\per\minute}) hence, it was used to create both the required gas current and to measure the radon concentration in the gas over time. 

\begin{center}
 \begin{minipage}{\textwidth}
  \begin{minipage}[b]{0.49\textwidth}
    \centering
  \includegraphics[width=\textwidth]{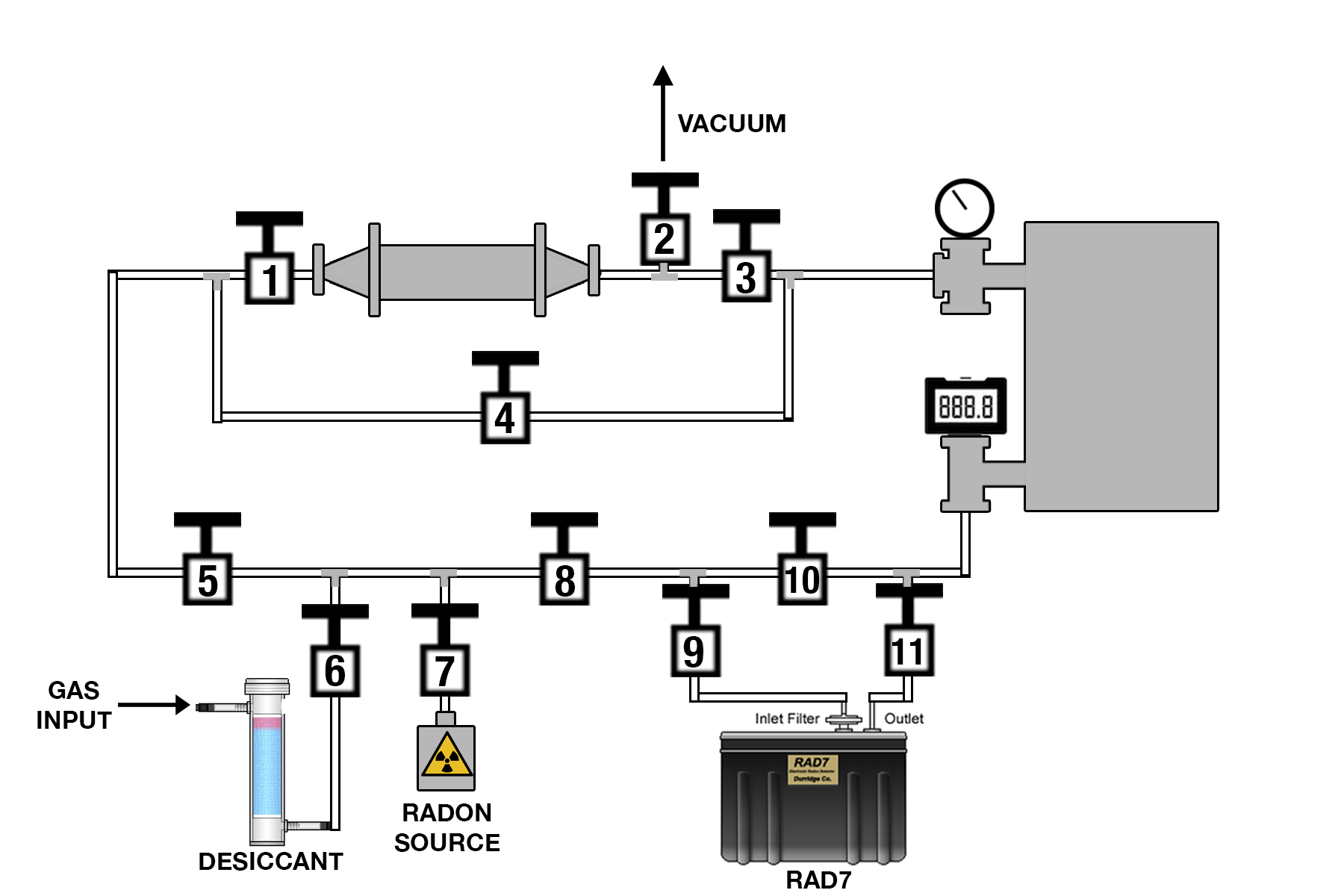} 
    \captionof{figure}{Schematic of the components used for the radon filtration from SF$_6$ tests. The valves are labelled with numbers.}
\label{radonfilt}
  \end{minipage}
  \hfill
  \begin{minipage}[b]{0.49\textwidth}
    \centering
  \begin{table}[H]
\centering

\scalebox{0.7}{
\begin{tabular}{|c|ccccccccccc|}
\hline
\multirow{2}{*}{\textbf{Step}}                                               & \multicolumn{11}{c|}{\textbf{Closed Valves}}                                                                                                                                                                                                                                                                                  \\
                                                                             & \textbf{1}                      & \textbf{2}           & \textbf{3}                      & \textbf{4}           & \textbf{5}           & \textbf{6}                      & \textbf{7}                      & \textbf{8}           & \textbf{9}                      & \textbf{10}          & \textbf{11}                      \\ \hline
Evacuation of System                                                         &                                 &                      &                                 &                      &                      & \textbullet                     & \textbullet                     &                      & \textbullet                     &                      & \textbullet                      \\ \hline
Input of Gas                                                                 &                                 & \textbullet          &                                 &                      &                      &                                 & \textbullet                     &                      & \textbullet                     &                      & \textbullet                      \\ \hline
\begin{tabular}[c]{@{}c@{}}Radon \\ Contamination\end{tabular}               & \textbullet                     & \textbullet          & \textbullet                     &                      &                      & \textbullet                     &                                 &                      &                                 & \textbullet          &                                  \\ \hline
\begin{tabular}[c]{@{}c@{}}Initial Concentration \\ Measurement\end{tabular} & \textbullet                     & \textbullet          & \textbullet                     &                      &                      & \textbullet                     & \textbullet                     &                      &                                 & \textbullet          &                                  \\ \hline
Engagement of Filter                                                         &                                 & \textbullet          &                                 & \textbullet          &                      & \textbullet                     & \textbullet                     &                      &                                 & \textbullet          &                                  \\ \hline
\begin{tabular}[c]{@{}c@{}}Filter Concentration \\ Measurement\end{tabular}  &                                 & \textbullet          &                                 & \textbullet          &                      & \textbullet                     & \textbullet                     &                      &                                 & \textbullet          &                                  \\ \hline
Changing of Filter                                                           & \multicolumn{1}{l}{\textbullet} & \multicolumn{1}{l}{} & \multicolumn{1}{l}{\textbullet} & \multicolumn{1}{l}{} & \multicolumn{1}{l}{} & \multicolumn{1}{l}{\textbullet} & \multicolumn{1}{l}{\textbullet} & \multicolumn{1}{l}{} & \multicolumn{1}{l}{\textbullet} & \multicolumn{1}{l}{} & \multicolumn{1}{l|}{\textbullet} \\ \hline
\end{tabular}}

\end{table}
     \captionof{table}{Steps of the radon filtration test operations and state of the valves. Closed valves are marked with solid black dots, unmarked valves were open in each of the operations. The numbers correspond to the valves in Figure \ref{radonfilt}.}
      \label{radvalves}
    \end{minipage}
  \end{minipage}
  
\end{center}
\vspace*{10mm}

\noindent The radon filtration tests were done as follows; the setup was first evacuated using a vacuum scroll pump for an hour and then filled with SF$_6$ gas via the desiccant to 1.1 bar, the operational pressure of the RAD7 detector. With the valves connecting the system to the filter closed, the internal air pump in the RAD7 was switched on to create the required gas current through the radon source so the gas becomes contaminated. The contamination process was done for an hour. After the contamination of the gas, the RAD7 was set to test for a total of 48 hours with a measurement recorded every hour. The first 24 hours was used to measure the initial concentration of radon. The filter was then engaged for the remaining 24 hours to measure the effect of the filter on the concentration of radon. The corresponding valves to close for each step of the test are shown in Table \ref{radvalves}.

The same set of gas was used throughout the measurements to avoid gas wastage. This led to different initial radon contamination concentration, depending on the amount of decay recorded before the subsequent experiments.

\subsection*{Radon Filtration Data Analysis}
The data recorded by the RAD7 include the number of decays per unit volume over time and relative humidity of the sample. Before the data was analysed, various corrections were required. This includes the use of the decay equation to model the recorded data. To do this, the initial radon concentration parameter was used to compare the effect of the molecular sieve filters to the absolute radon concentration. 

\subsubsection*{Corrections}
\paragraph*{Humidity} 
The $^{222}$Rn atom decays into a positively charged $^{218}$Po ion \cite{sutton1993}. The RAD7 uses this $^{218}$Po ion to measure the radon concentration. This is done in the internal hemisphere sample cell inside the detector, which has an electric field to drive the $^{218}$Po ion towards the detector \cite{durridge}. 
High humidity can be an issue because the slightly negative oxygen in water can attract the $^{218}$Po ion. A build up of water molecules around the $^{218}$Po ion can lead to the ion becoming neutralised hence preventing the ion from being detected inside the RAD7. Therefore, equation \ref{humidityeqn}, a correction for high humidity provided by Durridge, was applied \cite{durridge}. 

\begin{equation}
 C_{c}= \frac{100 C_{m}}{116.67-1.1\times RH} \,,
 \label{humidityeqn}
\end{equation}
where $C_{c}$ is the corrected radon concentration, $C_{m}$ is the measured radon concentration and RH is the relative humidity percentage for the RAD7 detector. 
This correction was applied to data points with relative humidity >15\si{\percent}. Overall, the average humidity for this measurement is 8\si{\percent}. This is mainly due to the remnant humidity and expected minuscule leak of the setup. 

\paragraph*{Radon Mixing Time}
The radon gas contaminate requires sufficient time to distribute evenly throughout the system. This is because during the contamination process some parts of the system may be more radon rich than others. This effect was evident in the data set as there was visible discontinuity between the data points before and after the mix. Therefore, the unmixed data were disregarded accordingly.

\subsubsection*{Fitting Data to the Decay Equation}

The decay relation shown in Equation \ref{decayeqn} was used to create a non linear regression fit to the data.
\begin{equation}
N(t)=N_0 e^{- \lambda t} \,,
\label{decayeqn}
\end{equation}

\noindent where $N(t)$ is the radon concentration at time t, $N_0$ is the initial radon concentration and $\lambda $ is the radon decay constant; calculated by using $^{222}$Rn half life of 3.8229 $\pm$ 0.00027  days \cite{martin1956}.

The data before and after the filter was engaged are considered separately. The sample size for the non linear regression fit to the \textit{filter on} data is adjusted until the discrepancy between the data and the decay equation is minimised. This is because the fit to the decay equation only applies when the filter is no longer active, for instance, when a new equilibrium is reached or the filter is saturated. The initial radon concentration was extrapolated from the best fit for both \textit{filter on} and \textit{filter off} data and used to compare the effective concentration change of radon caused by the engagement of the molecular sieve filter. 

\subsection*{Radon Filtration Results and Discussion}
\begin{figure}[h]
\begin{center}
\includegraphics[scale=0.2]{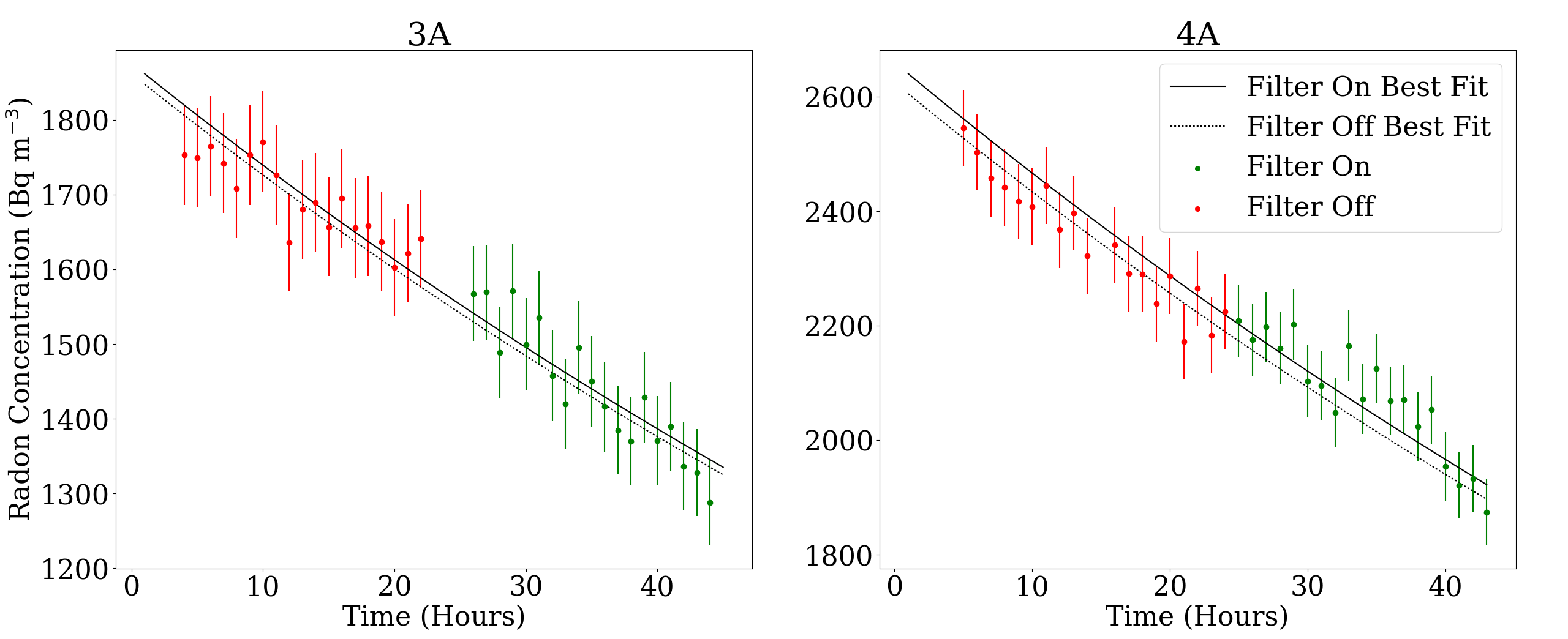}
\caption{Radon concentration shown as a function of time for the 3{\AA} (left) and 4{\AA} (right) molecular sieve filter tests. The filter was engaged after the first 24 hours. Data recorded before and after the filter was engaged are shown in red and green, respectively.}
\label{34A}
\end{center}
\end{figure}

\begin{figure}[h!]
\begin{center}
\includegraphics[scale=0.23]{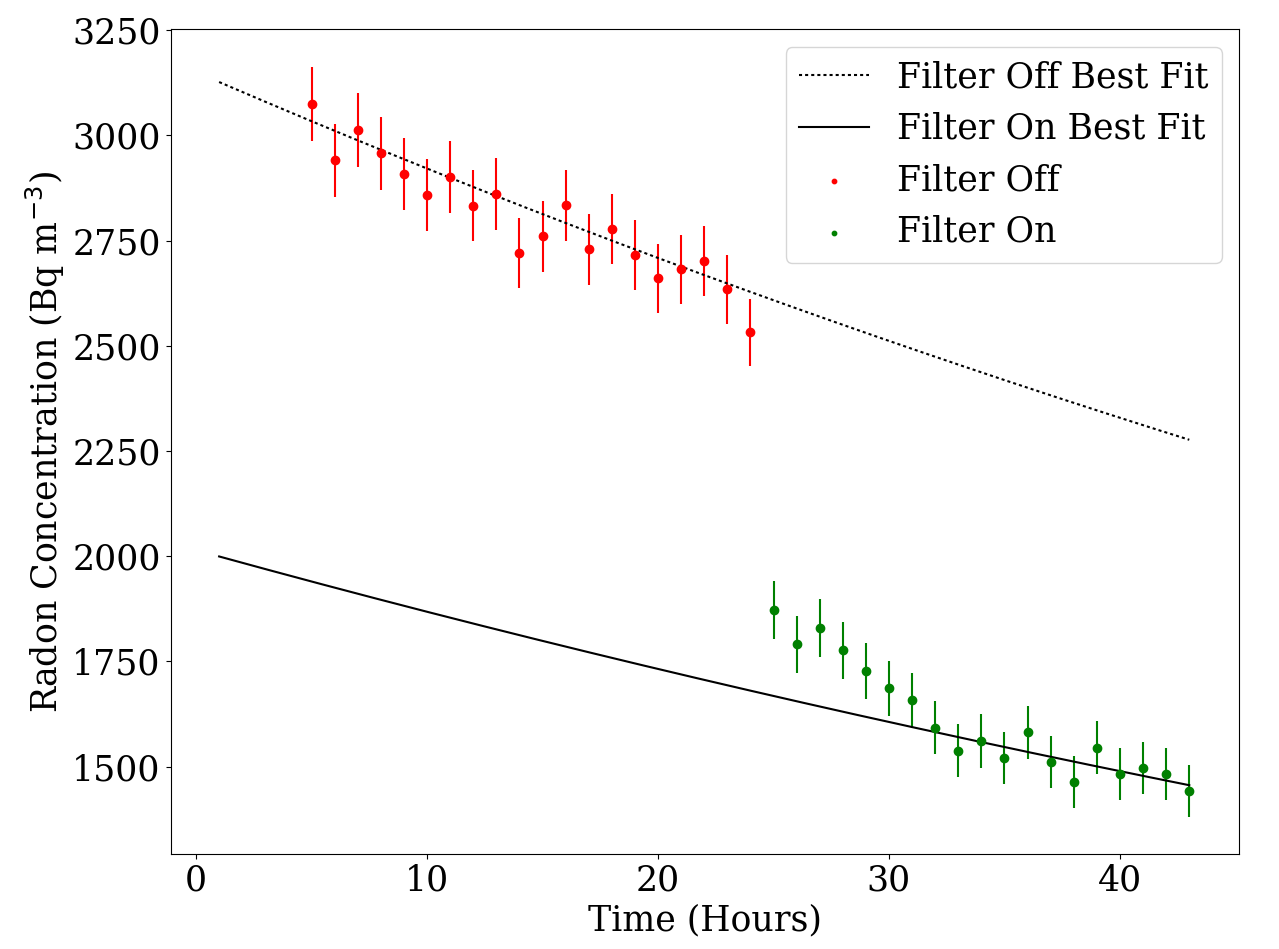}

\caption{Radon concentration shown as a function of time for the 5{\AA} molecular sieve filter. The filter was engaged after the first 24 hours.}

\label{5A}
\end{center}
\end{figure}

Results for the first radon filtration tests are shown in Figure \ref{34A}, Figure \ref{5A} and Table \ref{34a}. In these plots, the data points shown in red and green correspond to the measurements recorded in the first and second 24~hours  of  the test, when the filters were not engaged and engaged, respectively. Furthermore, the best fit to the decay equation are shown with a dotted and a solid line for the \textit{filter off} and \textit{filter on} data, respectively.

For the 3{\AA} molecular sieve test, it can be seen in Figure \ref{34A} that the radon concentration decay rate appears not to change when the 3{\AA} filter was engaged. This is also true for the 4{\AA} molecular sieve test as shown in Figure \ref{34A}.\\

\noindent The absolute change in concentration of radon caused by the introduction of a filter was determined by using the extrapolated initial radon concentration $N_0$. In Table \ref{34a}, for both the 3{\AA} and 4{\AA} filters, the extrapolated initial radon concentration $N_0$ for the \textit{filter on} and \textit{filter off} data are statistically consistent with theoretical predictions for the considered time window. This shows that the initial decay equation was obeyed even when the filter was engaged, making it evident that the engagement of the 3{\AA} or 4{\AA} filter has no overall effect on the concentration of radon.

\begin{table}[h]
\centering\scalebox{0.8}{
\begin{tabular}{|c|ccc|}
\hline
\multirow{2}{*}{\textbf{\begin{tabular}[c]{@{}c@{}}Molecular\\ Sieve\end{tabular}}} & \multicolumn{3}{c|}{\textbf{Extrapolated $N_0$} ( Bqm$^{-3}$)}                                                                                                                                                            \\ \cline{2-4} 
                                                                                    & \textbf{\begin{tabular}[c]{@{}c@{}}Filter Off\\ Data\end{tabular}} & \textbf{\begin{tabular}[c]{@{}c@{}}Filter On\\ Data\end{tabular}} & \textbf{\begin{tabular}[c]{@{}c@{}}On and Off\\ Data\end{tabular}} \\ \hline
3{\AA}                                                                                  & 1863.2 $\pm$ 9.1                                                    & 1875.6 $\pm$ 9.1                                                   & 1868.4 $\pm$ 6.9                                                    \\
4{\AA}                                                                                  & 2625.0 $\pm$ 11.1                                                    & 2657.2 $\pm$ 11.0                                                    & 2638.7 $\pm$ 11.1                                                    \\ \hline
\end{tabular}}
\caption{ Results from the extrapolated initial radon concentration parameter. The recorded \textit{filter on}, \textit{filter off} and the combined \textit{On and Off} data corresponds to the data points used when extrapolating the initial radon concentration. }
\label{34a}
\end{table}

\newpage 
\noindent The inability of the 3{\AA} and 4{\AA} molecular sieves in removing radon from SF$_6$ can be explained by the expected critical diameter of noble gases. For instance, the critical diameter of helium (argon) is 2{\AA} (3.8{\AA}) \cite{molecularsieves}. The location of these noble gases in the periodic table are in the first and third rows, respectively. Whereas radon is in the sixth row of the noble gas group. By considering the trend of atomic radii down a group, it is expected that the radon radii should be bigger than that of argon. Hence, it is expected that the critical diameter of radon is greater than the pores in the 3{\AA} and 4{\AA} molecular sieve filters.\\

\noindent Results for the 5{\AA} filter are shown in Figure \ref{5A}.  Here the decay rate is seen to significantly decrease after 24 hours. This time corresponds to when the gas was redirected through the 5{\AA} molecular sieve filter. This shows that the filter absorbs radon from SF$_6$.  A period of time after the significant drop in concentration, it appears that the rate of decay returns to obeying the decay rate equation.

\noindent To provide the best fit to the \textit{filter on} data the first 8 hours of data after the filter was engaged were not included to the decay equation fit.  The extrapolated initial radon concentration for the \textit{filter off} and \textit{filter on} data is 3150 $\pm$ 28 Bqm$^{-3}$ and 2014 $\pm$ 11 Bqm$^{-3}$, respectively. This results in a total initial radon concentration reduction of 36\%. The results show that it takes approximately 4~hours for the sieve to reach equilibrium with radon. It is expected that the rate of radon absorption in the molecular sieve can be affected by the operational pressure and flow rate of the gas \cite{kerry2007}, so these are not upper limit results. 

\section{Optimisation of Absorption Capabilities}
The behaviour of the absorption plot for the 5{\AA} molecular sieve seen in Figure \ref{5A} indicates that the filter saturates with time.  The number of radon atoms absorbed can easily be shown to be far smaller than the available number of pore holes.  So a more likely possibility is that an equilibrium situation is reached in which as many radon atoms are released as are absorbed.  A possible means to confirm this, and potentially increase the absorption, is to cool the radon sieve filter in an attempt to reduce the release of radon. 

\newpage
\subsection*{Application of Cold Trap }
\begin{figure}[h]
\begin{center}
\includegraphics[scale=0.15]{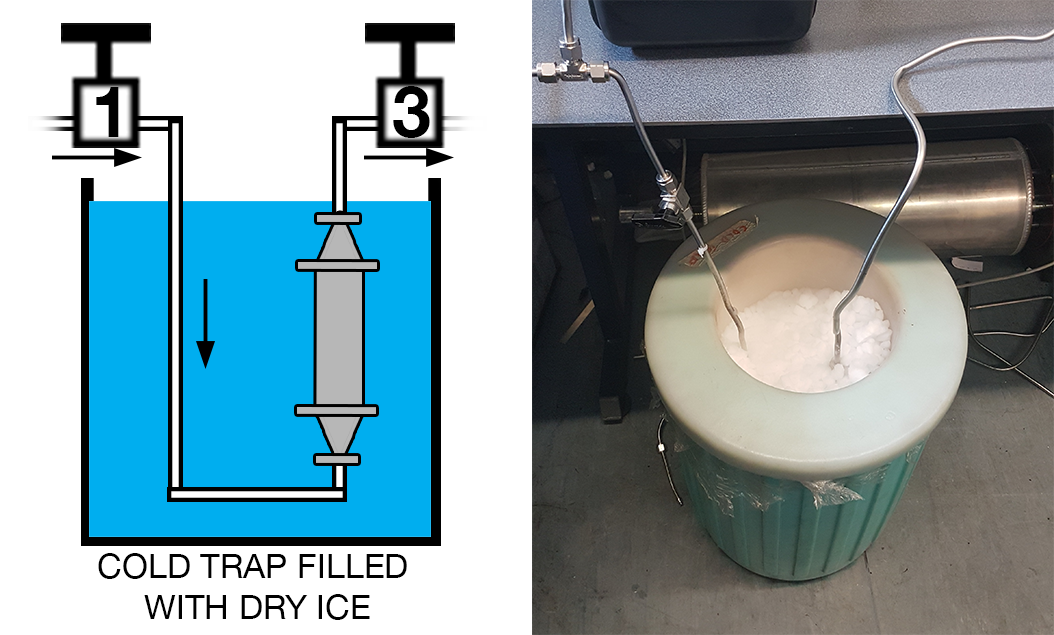} 
\caption{ Schematic of  the radon filtration setup in a cold trap and a photograph of the cold trap during a test.}
\label{trappic}
\end{center}
\end{figure}

The setup for testing the effect of cooling on the radon reduction capabilities of the 5{\AA} molecular sieve follows from the setup and procedure in Section \ref{sec:rad}. A small modification to the setup was made as shown in Figure \ref{trappic}; the stainless steel pipes between the molecular sieve and valve of the filter were extended to allow the filter to be placed inside a cold trap. Dry ice was used in the cold trap instead of liquid nitrogen as there was a concern about the O-ring's ability to operate at such low temperature. In this mode, the internal pump of the RAD7 radon detector was arranged such that the cold SF$_6$ flowed through the large emanation chamber before the radon concentration measurements. This was to ensure that the gas was within the RAD7 operating temperature.

 The total RAD7 testing time was extended to 50 hours. The first 20 hours was used to measure the initial radon concentration. After this, the molecular sieve filter was engaged and for the following 24 hours it was allowed to establish an equilibrium between adsorption and desorption of radon. For the remaining 6 hours the dry ice was added to the cold trap.

\subsection*{Cold Trap Results and Discussion}
\begin{figure}[h]
\begin{center}
\includegraphics[scale=0.3]{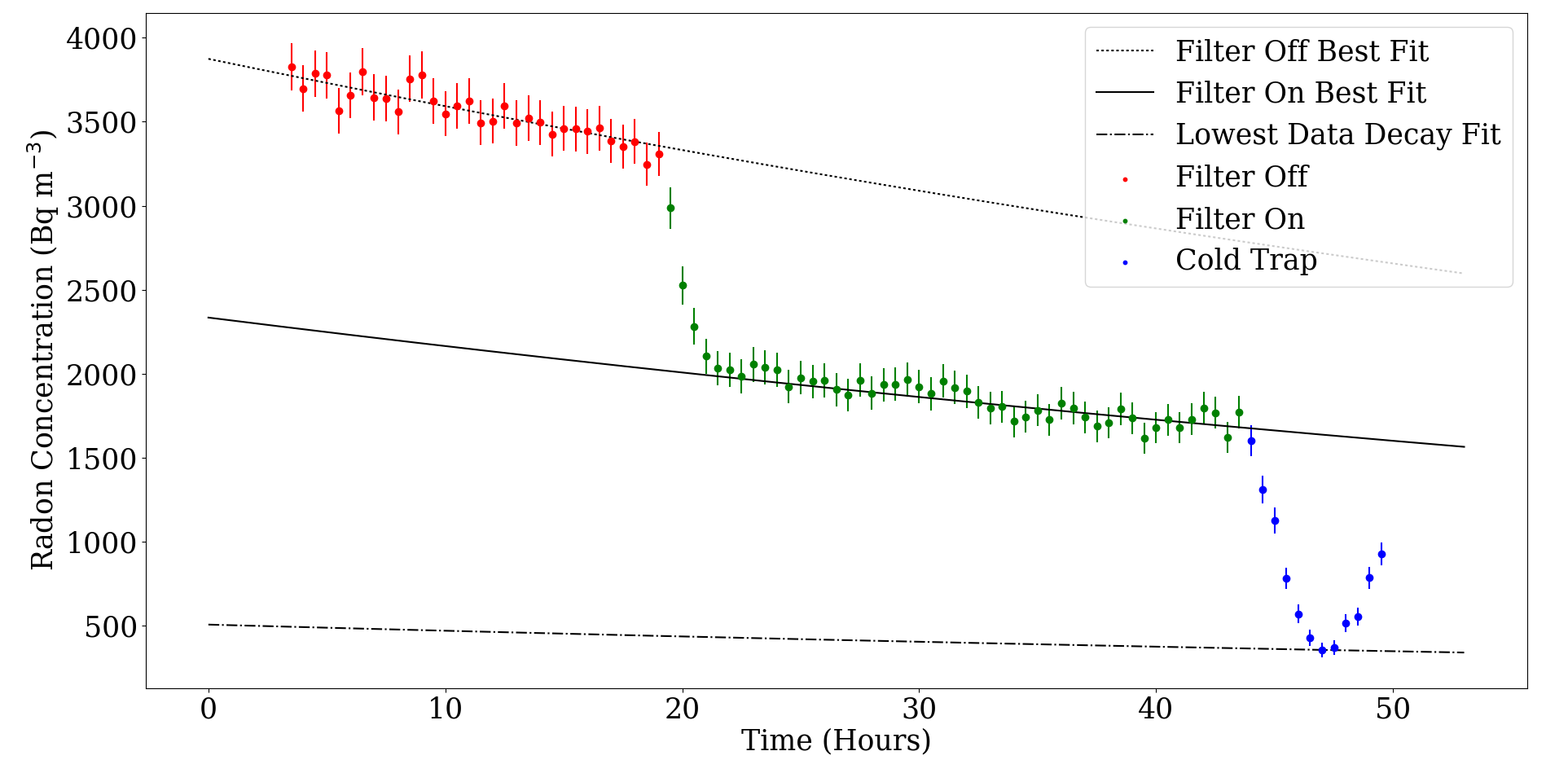} 
\caption{Radon concentration in SF$_6$ shown over time for the 5{\AA} molecular sieve filter. The filter was engaged after 20 hours and the cold trap was engaged after 44 hours. The decay fit on the blue data set was determined using only one data point to extrapolate the lowest possible radon concentration achieved.}
\label{trapresult}
\end{center}
\end{figure}

Results from the cooling test are shown in Figure \ref{trapresult}.  The analysis of the radon filtration data was performed using the same corrections and process to fit the decay as discussed in Section \ref{sec:rad}. In the radon concentration plot in Figure \ref{trapresult}, there are 3 sets of data points shown in red, green and blue. These data points correspond to when the filter was off, on and when the cold trap was engaged, respectively. There are also 3 radon decay equation fits; the \textit{filter off} best fit and \textit{filter on} best fit were computed using a non-linear regression model for the corresponding data. Whereas the decay fit for the lowest data point only used one data point  to extrapolate the lowest initial radon concentration achieved by the 5{\AA} molecular sieve with the cold trap.

As shown in Figure \ref{trapresult}, the initial radon concentration was reduced by the 5{\AA} filter and a new radon concentration equilibrium was reached as expected. Once the cold trap was introduced there was a further reduction in the concentration. This can be seen in the deviation of the cold trap data from the \textit{filter on} best fit. The lowest radon concentration was reached after 4 hours of introducing the cold trap. The increase in the radon concentration after this point is due to a gradual rise in temperature as the dry ice started to evaporate.\\

\noindent The extrapolated initial radon concentration for the decay fits are shown in Table \ref{coldtraptable}. This allows for comparison of the effective radon reduction of the 5{\AA} molecular sieve filter and the 5{\AA} molecular sieve filter with the cold trap. Results show that the application of the 5{\AA} molecular sieve with the cold trap reduced the initial radon concentration by a total of 87\%.

\begin{table}[h]
\centering
\scalebox{0.7}{
\begin{tabular}{|c|c|c|}
\hline
\textbf{Data}             & \textbf{\begin{tabular}[c]{@{}c@{}}Extrapolated $N_0$ \\ ( Bqm$^{-3}$)\end{tabular}} & \textbf{\begin{tabular}[c]{@{}c@{}}Total Radon \\ Concentration Reduction\end{tabular}} \\ \hline
\textbf{Filter Off}       & 3874.8 $\pm$ 13.1                                                                       & -                                                                                       \\
\textbf{Filter On}        & 2356.9 $\pm$ 10.0                                                                       & 40\%                                                                                    \\
\textbf{Cold Trap Lowest} & 504.6                                                                                 & 87\%                                                                                    \\ \hline
\end{tabular}}
\caption{Results from the extrapolated initial radon concentration parameter for the cold trap test. The total radon concentration reduction is the percentage reduction with respect to the initial radon contamination concentration.}
\label{coldtraptable}
\end{table}

\section{Conclusions}
In this work four types of molecular sieves, 3{\AA}, 4{\AA}, 5{\AA} and 13X from Sigma-Aldrich, were used to investigate the radon absorption capabilities from SF$_6$ gas. It was found that the 13X type absorbed SF$_6$ molecules, whereas the 3{\AA}, 4{\AA} and 5{\AA} types did not. The 5{\AA} molecular sieve was the only molecular sieve type that successfully reduced the radon concentration. This achieved a reduction in radon concentration from 3150 $\pm$ 28 Bqm$^{-3}$ to 2014 $\pm$ 11 Bqm$^{-3}$ which is approximately 36.1\% of the initial radon concentration in the contaminated SF$_6$. It was found that the molecular sieves were not saturated by the radon atoms, and so the 5{\AA} radon reduction capabilities were further optimised by applying a cold trap containing dry ice. The combination of the 5{\AA} molecular sieve and a cold trap resulted in reduction of the radon concentration to approximately 87.0\% of the initial radon concentration. The ability of the 5{\AA} molecular sieve filter with a cold trap to significantly reduce radon concentration from SF$_6$ provides a promising foundation for the construction of a radon filtration set up for future ultra-sensitive SF$_6$ gas rare-event physics experiments. It should be noted that the 5{\AA} molecular sieve structure contains a form of aluminium which could potentially be a source of low-level radon radioactivity.  However, this heavily depends of the purity of the starting materials and method of synthesis used by the manufacturer. The ability of a given mass of the sieve to reach the required $\sim$\si{\micro\becquerel\per\litre} radon level for a typical dark matter experiment depends on the initial radon concentration of the target SF$_6$ gas. Higher gas purity can be achieved by increasing the number of available 5{\AA} molecular sieve pores.  At least one dark matter group to our knowledge, XMASS, has developed an ultra-low radon emitting molecular sieve \cite{lee2016}.

\acknowledgments
The authors would like to acknowledge support for this work through the STFC grant award ST/P00573X/1 and associated awards from STFC.

\end{document}